\documentclass{webofc}\usepackage[varg]{txfonts}
\usepackage{bm}\usepackage{color}\usepackage{booktabs}

\begin{document}\title{Multiquark-Adequate QCD Sum Rules:\\the Case
of Flavour-Exotic Tetraquarks}\author{Wolfgang Lucha\inst{1}\fnsep
\thanks{\email{Wolfgang.Lucha@oeaw.ac.at}}\and Dmitri Melikhov
\inst{1,2,3}\fnsep\thanks{\email{dmitri_melikhov@gmx.de}}\and
Hagop Sazdjian\inst{4}\fnsep\thanks{\email{sazdjian@ipno.in2p3.fr}
}}\institute{Institute for High Energy Physics, Austrian Academy
of Sciences, Nikolsdorfergasse 18,\\A-1050 Vienna, Austria\and
D.~V.~Skobeltsyn Institute of Nuclear Physics, M.~V.~Lomonosov
Moscow State University,\\119991 Moscow, Russia\and Faculty of
Physics, University of Vienna, Boltzmanngasse 5, A-1090 Vienna,
Austria\and Institut de Physique Nucl\'eaire, Universit\'e
Paris-Sud, CNRS-IN2P3, Universit\'e Paris-Saclay,\\91405 Orsay,
France}

\abstract{Frequently, theoretical discussions of multiquark hadron
states prove to be contaminated or even dominated by contributions
of conventional hadrons. For the approach to QCD bound states in
terms of QCD sum rules,~we~show~how to get rid of all unwanted
ordinary-hadron ballast by boiling down the traditional QCD
sum-rule formalism to the nonconventional aspects in the focus of
interest.}\maketitle

\section{Incentive and Findings: Rigorous Adequacy of Multiquark
Studies}The notion of \emph{multiquark hadrons\/} comprises each
bound state composed of a larger number of quarks than any
ordinary quark--antiquark meson or ordinary three-quark baryon
calls~its own. In view of the slowly but steadily increasing
evidence for the actual experimental observation, and consequently
unambiguous confirmation of the existence, of tetraquarks and
pentaquarks, we recently embarked on a rigorous theoretical
investigation of multiquarks
\cite{TQN1,TQP1,TQN2,TQP2,ESR,TAQSR,CROMO,ESR2}. A standard tool
for an extraction of properties of hadronic bound states from the
underlying quantum field theory, quantum chromodynamics (QCD), is
offered by a technique called QCD sum~rules~\cite{QSR}.

QCD sum rules represent a nonperturbative approach to bound states
of quarks and gluons (the degrees of freedom of QCD) in form of
analytic relationships between observable hadron properties and
the parameters of QCD (strong coupling and quark masses). They can
be found upon evaluation of correlation functions of hadron
interpolating operators, defined in terms of quark and gluon
fields, at both phenomenological, i.e., hadronic, and QCD levels,
by insertion of complete sets of hadron states, conversion of
nonlocal operator products into series of local operators by
exploiting the operator product expansion \cite{KGW}, removal of
subtraction terms and suppression of hadron contributions above
ground states by applying Borel transformations to new variables,
and mutual cancellation of hadronic and perturbative QCD
contributions above Borel-variable-dependent effective thresholds
\cite{LMST1,LMST2,LMST3}. The final outcome comprises, at QCD
level, both purely perturbative contributions (conveniently
represented by dispersion integrals of spectral densities) and
nonperturbative contributions, parametrized by vacuum condensates
and nicknamed ``power corrections'' (since getting multiplied by
powers of Borel~parameters).\pagebreak

Given the above general recipe, it is, in principle,
straightforward to derive QCD sum rules from $n$-point Green
functions of (appropriately defined) hadron interpolating
operators. In the case of all multiquark hadrons, however, due
care has to be paid to the proper disentanglement of those
contributions to any such correlation function that may involve
the kind of multiquark states of interest and those contributions
that are definitely not related to the states under study since,
for instance, they affect exclusively ordinary hadrons
\cite{ESR,TAQSR,ESR2}. More precisely, from our point of view all
the latter contributions should be unambiguously identified,
stripped~off from any QCD sum rule found in the first step by
na\"ive application of the formalism, and~discarded.

In the following, for the sake of illustration, let us focus to
the particular case of \emph{tetraquark mesons\/} contributing in
form of intermediate states to some scattering of two ordinary
mesons. Quite generally, indices $a,b,c,\dots$ will label the
\emph{flavour\/} quantum numbers of the (anti)quarks.

Any contribution of the first kind, i.e., presumably capable of
providing information about tetraquarks in the focus of interest,
is termed \emph{tetraquark-phile\/} \cite{TQP1,TQP2}: retaining
exclusively that sort of contributions entails QCD sum rules
enjoying the wanted tetraquark adequacy \cite{ESR,TAQSR,ESR2}.
Phrased, a little bit more technically, in terms of Feynman
diagrams, in order to be regarded as tetraquark-phile a Feynman
diagram should depend on the appropriate Mandelstam variable $s$
in a non-polynomial manner and must develop a branch cut starting
at a branch point $\hat s$ defined by the square of the sum of the
masses $m_a$, $m_b$, $m_c$, $m_d$ of all involved (anti)quarks
$\overline q_a$, $q_b$, $\overline q_c$, $q_d$ acting as
constituents of the envisaged tetraquark bound state \cite{TQN1}:
$\hat s=(m_a+m_b+m_c+m_d)^2.$ The presence or absence of such cuts
is easily verified by application of the Landau equations
\cite{LDL}. Thus established cuts betray the diagram's support of
adequate four-quark intermediate~states. The latter, in turn, may
contribute to the development of a pole related to a physical
tetraquark.

Mutatis mutandis, our concepts \cite{ESR} of sharpening the
traditional QCD sum-rule formalism apply, rather generally, to
arbitrary kinds of multiquark hadrons. Hence merely for the sake
of definiteness, here we outline our lines of argument for the
particular case of tetraquark~mesons of genuinely exotic
quark-flavour content, i.e., bound states $T$ of two antiquarks
$\overline q_a$,~$\overline q_c$ and~two quarks $q_b$, $q_d$ the
flavour quantum numbers $a,b,c,d$ of which are definitely mutually
different:\begin{equation}T=[\overline q_a\,q_b\,\overline q_c\,
q_d]\ ,\qquad a,b,c,d\in\{u,d,s,c,b\}\ .\label{T}\end{equation}

The starting point of the derivation of some QCD sum rule is the
choice of a convenient set of interpolating operators. This
endeavour is non-negligibly facilitated by the observation that
(upon application of adequate Fierz transformations) any
conceivable tetraquark interpolating operator can be shown to be
equivalent to a sum of products of colour-singlet quark--antiquark
bilinear operators. For our study, the spinor structure of any
encountered operator is irrelevant and thus notationally
suppressed. Then, our basic operators will be the quark-bilinear
currents\begin{equation}j_{\overline ab}(x)\equiv\overline q_a(x)
\,q_b(x)\ .\label{b}\end{equation}For any fully flavour-exotic
tetraquark (\ref{T}), only two interpolating currents can be
constructed:\begin{equation}\theta_{\overline ab\overline cd}(x)
\equiv j_{\overline ab}(x)\,j_{\overline cd}(x)\ ,\qquad
\theta_{\overline ad\overline cb}(x)\equiv j_{\overline ad}(x)\,
j_{\overline cb}(x)\ .\label{c}\end{equation}

Our ultimate goal is the extraction of predictions, emerging from
QCD, for the elementary features of the tetraquark (\ref{T}), that
is, its mass $M$, characterizing the location of its pole, its two
decay constants $f_{\overline ab\overline cd}$ and $f_{\overline
ad\overline cb}$, defined in terms of the two tetraquark currents
(\ref{c}) according~to$$f_{\overline ab\overline cd}\equiv
\langle0|\theta_{\overline ab\overline cd}|T\rangle\ ,\qquad
f_{\overline ad\overline cb}\equiv\langle0|\theta_{\overline
ad\overline cb}|T\rangle\ ,$$as well as its transition amplitudes
to quark-flavour-matching two-conventional-meson states.

To this end, it proves advantageous, from the technical point of
view, to start such analyses from four-point Green functions of
ordinary-meson interpolating operators, i.e., from vacuum
expectation values of time-ordered (T) products of four
quark-bilinear currents of the kind (\ref{b}),\pagebreak
\begin{equation}\left\langle{\rm T}\left(j(y)\,j(y')\,j^\dag(x)\,
j^\dag(x')\right)\right\rangle.\label{G}\end{equation}From these
four-point Green functions, all correlation functions we are
interested in then arise via configuration-space pair contractions
of quark-bilinear currents (\ref{b}), namely, the two-point
correlator of two tetraquark interpolating currents (\ref{c}) from
the contraction of two such $j$ pairs,\begin{equation}\left\langle
{\rm T}\left(\theta(y)\,\theta^\dag(x)\right)\right\rangle=
\lim_{\underset{\scriptstyle y'\to y}{\scriptstyle x'\to x}}
\left\langle{\rm T}\left(j(y)\,j(y')\,j^\dag(x)\,j^\dag(x')\right)
\right\rangle,\label{2p}\end{equation}the three-point correlator
of one tetraquark and two $j$ currents by just a single pair
contraction,\begin{equation}\left\langle{\rm T}\left(j(y)\,j(y')\,
\theta^\dag(x)\right)\right\rangle=\lim_{x'\to x}\left\langle{\rm
T}\left(j(y)\,j(y')\,j^\dag(x)\,j^\dag(x')\right)\right\rangle.
\label{3p}\end{equation}

Dealing with flavour-exotic tetraquarks, we must discriminate
between two \emph{quark-flavour topologies\/} characterized by
whether the quark-flavour distributions in initial and final state
are the same or different and, unsurprisingly, analyze separately
\cite{ESR,TAQSR,ESR2} two disjoint categories of processes: those
that do not undergo quark rearrangement (Sect.~\ref{4p}) and those
that do (Sect.~\ref{4r}), called the flavour-preserving (vulgo
``direct'') and -reshuffling (vulgo ``recombination'') cases.

\begin{figure}[b]\centering\includegraphics[scale=.398841,clip]
{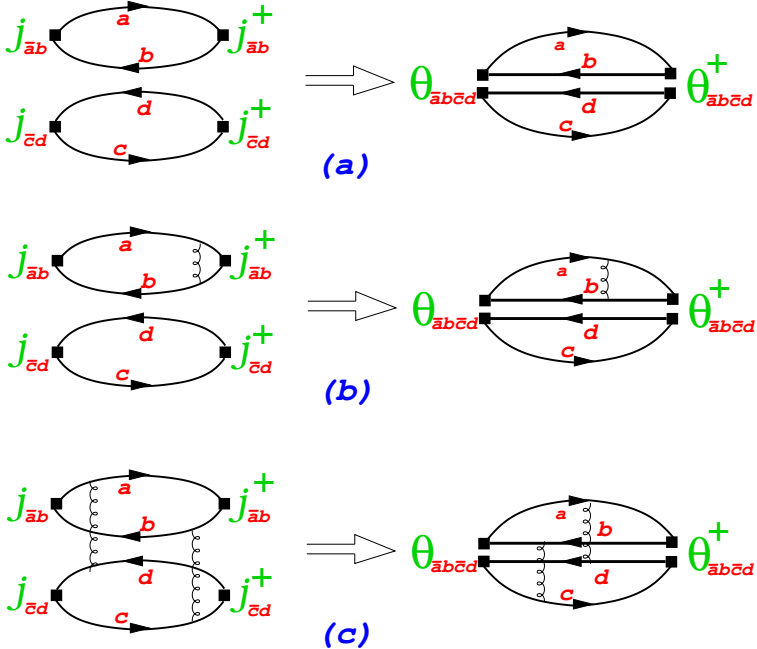}\caption{Typical examples of contributions at
the order $O(\alpha_{\rm s}^0)$ (a), $O(\alpha_{\rm s})$ (b), and
$O(\alpha_{\rm s}^2)$ (c) of the strong coupling $\alpha_{\rm s}$
(corresponding to the internal exchanges of no, one or two gluons,
respectively, indicated by curly lines) to a flavour-preserving
Green function $\left\langle{\rm T}\left(j_{\overline ab}\,
j_{\overline cd}\,j^\dag_{\overline ab}\,j^\dag_{\overline cd}
\right)\right\rangle$ of four quark-bilinear currents $j^{(\dag)}$
(left), and the contraction to the correlator $\left\langle{\rm
T}\left(\theta_{\overline ab\overline cd}\,\theta^\dag_{\overline
ab\overline cd}\right)\right\rangle$ of two tetraquark operators
$\theta^{(\dag)}_{\overline ab\overline cd}$
(right)~\cite{ESR,TAQSR}.}\label{FP}\end{figure}We arrive at an
identical conclusion \cite{ESR,TAQSR,ESR2}: Within a perturbative
expansion in powers~of$$\alpha_{\rm s}\equiv\frac{g_{\rm
s}^2}{4\pi}$$(with the strong coupling $g_{\rm s}$, together with
all quark masses, constituting the set of parameters of QCD), the
tetraquark-phile Feynman diagrams are necessarily of order
$O(\alpha_{\rm s}^2)$ or higher \pagebreak and must involve, at
least, two gluon exchanges of appropriate \emph{non-separable\/}
topology, in order to exhibit the requisite four-quark cut in the
Mandelstam variable $s$. Consequent implementation of
quark--hadron duality renders possible to systematically reject
all unambiguously identified ordinary-hadron contamination. Only
multiquark-adequate QCD sum rules incorporating and reflecting
these insights should be regarded as providing reliable
descriptions of multiquarks.\footnote{For pentaquarks too,
problems attributable to a too na\"ive application of QCD sum
rules have been noted \cite{KMN,NKMKE}.}

\section{Four-Quark Processes Retaining the Distribution of Quark
Flavour}\label{4p}
\begin{figure}[h]\centering\includegraphics[scale=.27765,clip]
{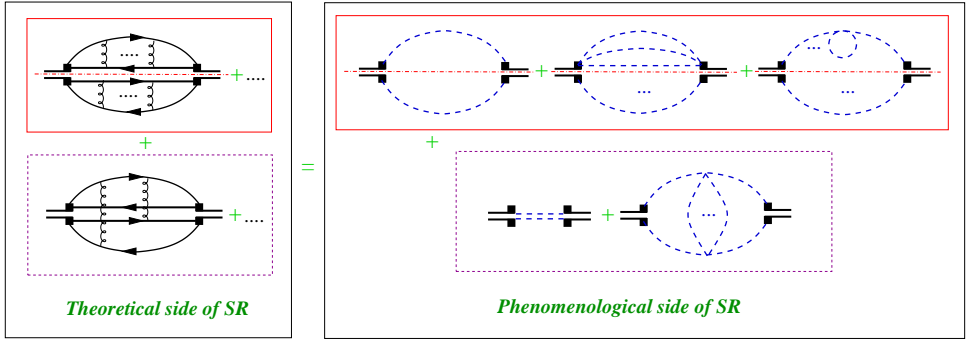}\caption{Disentanglement of the QCD sum rule (SR) for
any generic \emph{flavour-preserving\/} Green function
$\left\langle{\rm T}\left(\theta_{\overline ab\overline cd}\,
\theta^\dag_{\overline ab\overline cd}\right)\right\rangle$ of
two, by request flavour-exotic, tetraquark operators $\theta
^{(\dag)}_{\overline ab\overline cd}$ into a pair of QCD sum rules
(indicatively separated by a red horizontal dot-dashed line) for
the Green functions of two quark-bilinear currents $j^{(\dag)}$
interpolating ordinary mesons (generically indicated by blue
dashed lines) (top),~and a QCD sum rule for the
\emph{non-separable\/} part of such Green function $\left\langle
{\rm T}\left(\theta_{\overline ab\overline cd}\,\theta^\dag
_{\overline ab\overline cd}\right)\right\rangle$ (bottom). Only
the latter can develop a tetraquark pole (blue dashed double line)
exhibiting the $\theta_{\overline ab\overline cd}$ flavour degrees
of freedom~\cite{ESR,TAQSR}.}\label{ESR}\end{figure}

For four-point Green functions (\ref{G}) of
quark-bilinear operators (\ref{b}) with \emph{identical\/}
quark-flavour structure of initial and final states, the proof of
our claim is, in fact, rather straightforward, and readily
exemplified for the two-point correlators (\ref{2p}) emerging from
their contractions~(Fig.~\ref{FP}),$$\left\langle{\rm T}
\left(\theta_{\overline ab\overline cd}(y)\,\theta^\dag_{\overline
ab\overline cd}(x)\right)\right\rangle,\qquad\left\langle{\rm T}
\left(\theta_{\overline ad\overline cb}(y)\,\theta^\dag_{\overline
ad\overline cb}(x)\right)\right\rangle.$$Feynman graphs of orders
$O(\alpha_{\rm s}^0)$ and (partly due to vanishing colour traces)
$O(\alpha_{\rm s})$ involve~two \emph{unconnected\/} quark loops:
each contributes to the traditional QCD sum rule of its own
(Fig.~\ref{ESR}).

\begin{figure}[h]\centering\includegraphics[scale=.47736,clip]
{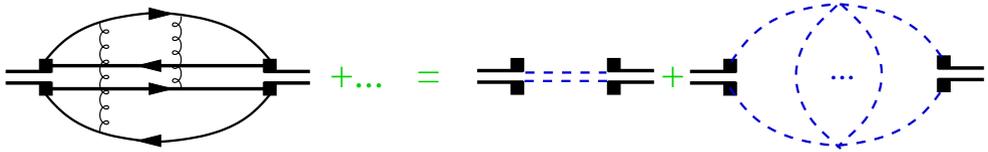}\caption{Tetraquark-adequate QCD sum rule for
each flavour-retaining Green function $\left\langle{\rm T}\left(
\theta_{\overline ab\overline cd}\,\theta^\dag_{\overline
ab\overline cd}\right)\right\rangle$, relating tetraquark-phile
Feynman-diagram contributions [involving at least two gluon
exchanges (curly lines), thus of order $O(\alpha_{\rm s}^2)$ or
higher] at the QCD level to contributions of potential tetraquark
poles (blue dashed double line) and just non-separable meson
diagrams (blue dashed lines) at the hadron level
\cite{ESR,TAQSR}.}\label{ESRf}\end{figure}\noindent Thus,
tetraquark-phile contributions are, at least, of order
$O(\alpha_{\rm s}^2)$. Only these (Fig.~\ref{ESRf}) should be
retained in the construction of QCD sum rules assumed to describe
\emph{non-conventional\/} hadrons.

\begin{figure}[t]\centering\includegraphics[scale=.40108,clip]
{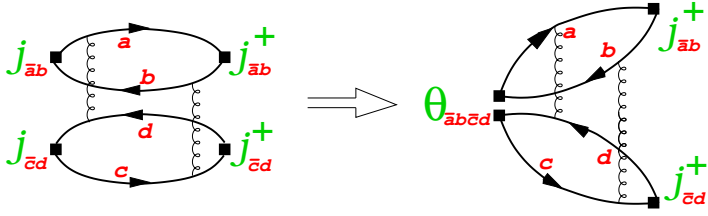}\caption{Example of a contribution of
lowest \emph{tetraquark-phile\/} strong-coupling order
$O(\alpha_{\rm s}^2)$, enabled~by exchange of two gluons (curly
lines), to the flavour-preserving Green function $\left\langle{\rm
T}\left(j_{\overline ab}\,j_{\overline cd}\,j^\dag_{\overline ab}
\,j^\dag_{\overline cd}\right)\right\rangle$ of four
quark-bilinear currents $j^{(\dag)}$ (left), contracted to a
correlator $\left\langle{\rm T}\left(j_{\overline ab}\,
j_{\overline cd}\,\theta^\dag_{\overline ab\overline cd}\right)
\right\rangle$ of a tetraquark current $\theta^\dag_{\overline ab
\overline cd}$ and two quark-bilinear currents $j_{\overline ab}$
and $j_{\overline cd}$ (right) underlying the
tetraquark--two-meson transition \cite{ESR,TAQSR}.}\label{3P}
\end{figure}\noindent The same holds, by its construction
(\ref{3p}), for the flavour-retaining (Fig.~\ref{3P}) three-point
correlator$$\left\langle{\rm T}\left(j_{\overline ab}(y)\,
j_{\overline cd}(y')\,\theta^\dag_{\overline ab\overline cd}(x)
\right)\right\rangle.$$

If stripping off all conventional contributions taken care of by
QCD sum rules for \emph{ordinary\/} mesons, we are left with
flavour-\underline{p}reserving \emph{tetraquark-adequate\/} QCD
sum rules involving, in their dispersion representation, solely
tetraquark-phile spectral densities $\rho_{\rm p},\Delta_{\rm p}$
and the related power corrections, as well as Borel parameters
$\tau$, $\tau$-dependent effective thresholds $s_{\rm eff}$, and,
in the case of three-point functions, Fourier transforms $A(T\to
j_{\overline ab}\,j_{\overline cd})$ of $\langle0|{\rm
T}[j_{\overline ab}(y)\,j_{\overline cd}(y')]|T\rangle$:

\begin{figure}[b]\centering\includegraphics[scale=.40108,clip]
{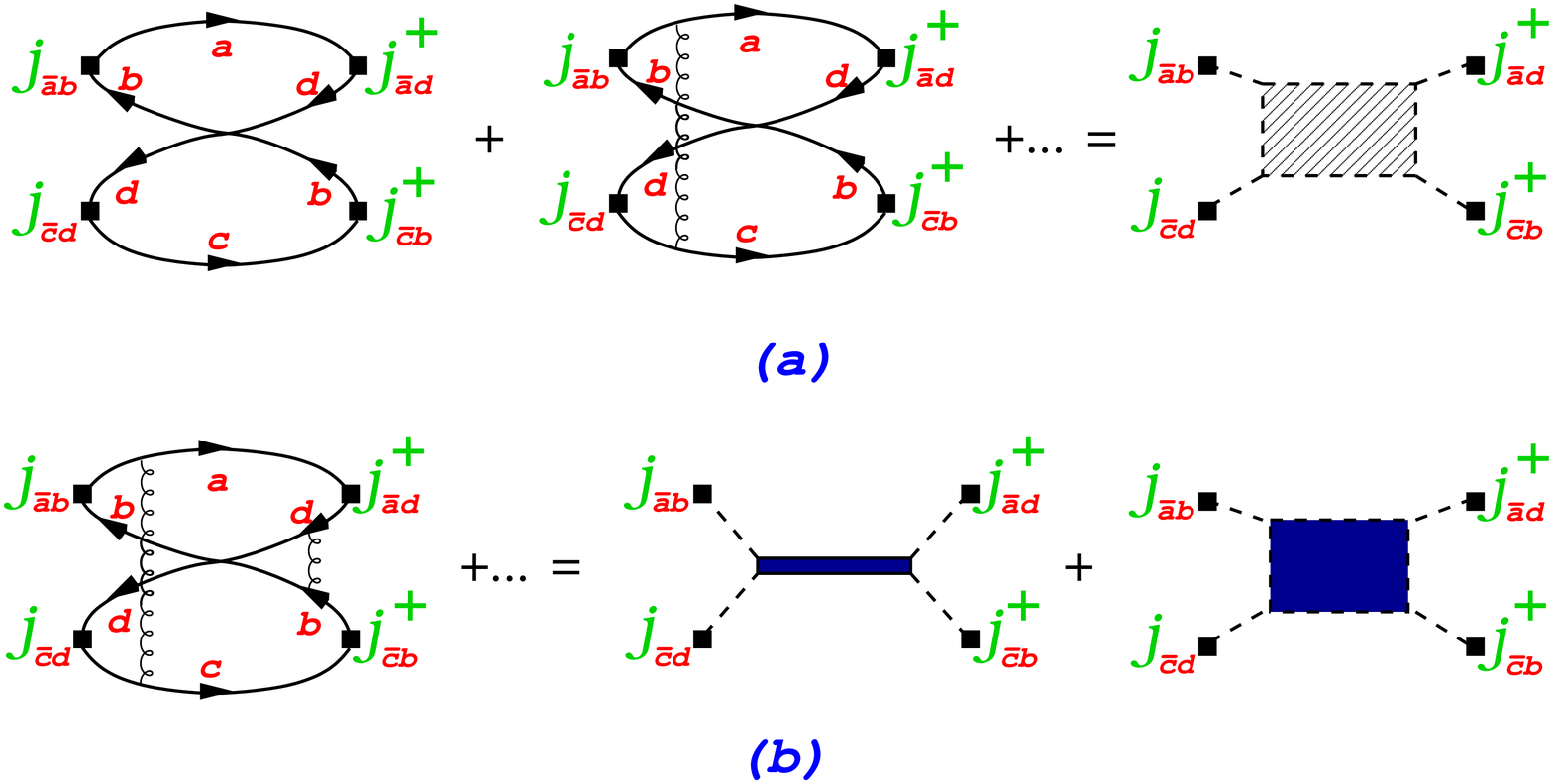}\caption{Disentanglement of the
contributions to a flavour-rearranging correlator $\left\langle
{\rm T}\left(j_{\overline ad}\,j_{\overline cb}\,j^\dag_{\overline
ab}\,j^\dag_{\overline cd}\right)\right\rangle$ of four
\emph{flavour-permuting\/} quark-bilinear operators $j^{(\dag)}$
into (a) QCD-level contributions \emph{without\/} four-quark
$s$-channel cuts (left), mirrored by hadron-level contributions
\emph{without} two-meson $s$-channel cuts (hatched rectangle,
right), and (b) tetraquark-phile QCD-level contributions
developing four-quark $s$-channel cuts (left), hence dual to
hadron-level contributions (right) encompassing possible
tetraquark $s$-channel poles (blue horizontal bar) and
contributions comprising a two-meson $s$-channel cut
(blue~filled~rectangle) \cite{ESR2}.}\label{ESRR}\end{figure}
\begin{align*}(f_{\overline ab\overline cd})^2\exp(-M^2\,\tau)&=
\hspace{-5.6ex}\int\limits_{(m_a+m_b+m_c+m_d)^2}^{s_{\rm eff}
(\tau)}\hspace{-5.1ex}{\rm d}s\exp(-s\,\tau)\,\rho_{\rm p}(s)+
\mbox{Borelized power corrections}\ ,\\[1ex]f_{\overline
ab\overline cd}\,A(T\to j_{\overline ab}\,j_{\overline cd})
\exp(-M^2\,\tau)&=\hspace{-5.6ex}
\int\limits_{(m_a+m_b+m_c+m_d)^2}^{s_{\rm eff}(\tau)}
\hspace{-5.1ex}{\rm d}s\exp(-s\,\tau)\,\Delta_{\rm p}(s)+
\mbox{Borelized power corrections}\ .\end{align*}

\section{Four-Quark Reactions Rearranging the Quark-Flavour
Distribution}\label{4r}
\begin{figure}[h]\centering\includegraphics[scale=.40108,clip]
{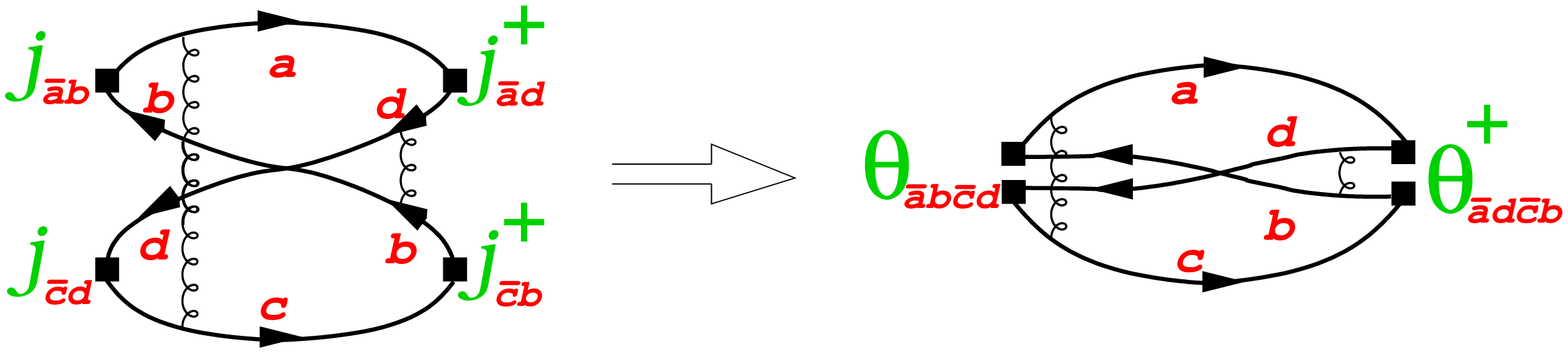}\caption{Example of a tetraquark-phile, lowest
strong-coupling order-$O(\alpha_{\rm s}^2)$, two-gluon-enabled
(curly lines) contribution to \emph{flavour-permuted\/}
correlators $\left\langle {\rm T}\left(j_{\overline ad}\,
j_{\overline cb}\,j^\dag_{\overline ab}\,j^\dag_{\overline cd}
\right)\right\rangle$ of four quark-bilinear currents $j^{(\dag)}$
(left), contracted to correlators $\left\langle{\rm T}\left(\theta
_{\overline ad\overline cb}\,\theta^\dag_{\overline ab\overline
cd}\right)\right\rangle$ of two tetraquark operators $\theta_
{\overline ad\overline cb}$ and $\theta^\dag_{\overline ab
\overline cd}$ (right)~\cite{ESR,TAQSR,ESR2}.}\label{2R}
\end{figure}

\begin{figure}[h]\centering\includegraphics[scale=.40108,clip]
{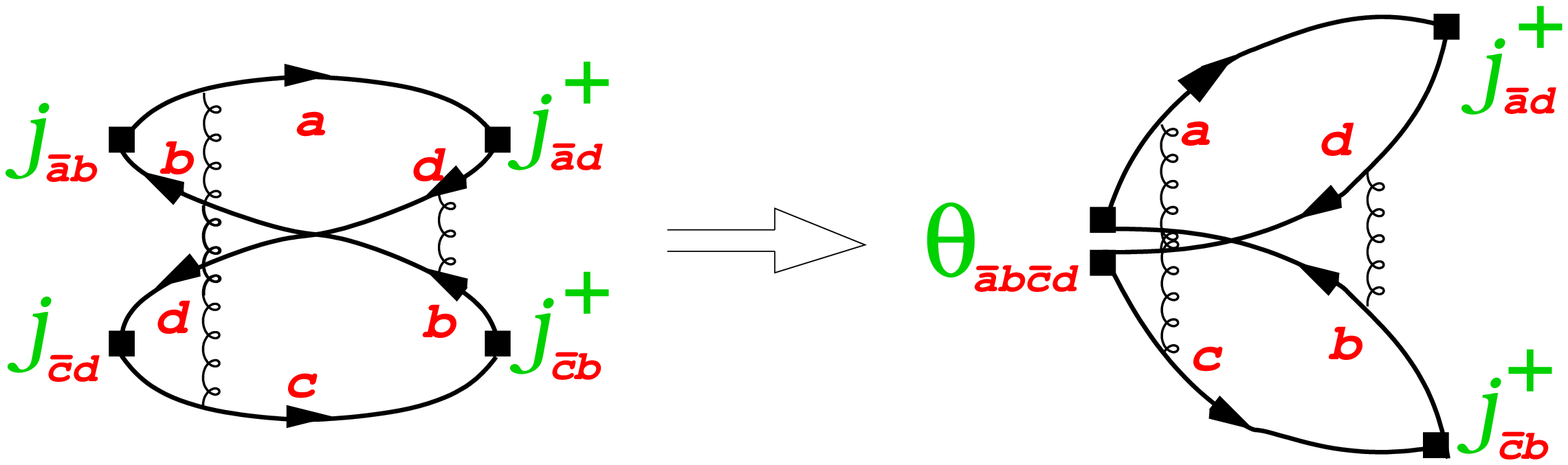}\caption{Example of a contribution of
lowest \emph{tetraquark-phile\/} strong-coupling order
$O(\alpha_{\rm s}^2)$, enabled~by exchange of two gluons (curly
lines), to a \emph{flavour-rearranging\/} Green function
$\left\langle{\rm T}\left(j_{\overline ad}\,j_{\overline cb}\,
j^\dag_{\overline ab}\,j^\dag_{\overline cd}\right)\right\rangle$
of four quark-bilinear currents $j^{(\dag)}$ (left), contracted to
a correlator $\left\langle{\rm T}\left( j_{\overline ad}\,j_
{\overline cb}\,\theta^\dag_{\overline ab \overline cd}\right)
\right\rangle$ of a tetraquark current $\theta^\dag_{\overline
ab\overline cd}$ and two quark-bilinear currents $j_{\overline
ad}$ and $j_{\overline cb}$ (right) underlying the
tetraquark--two-meson transition \cite{TAQSR,ESR2}.}\label{3R}
\end{figure}

For a four-point Green function (\ref{G})
of quark-bilinear currents (\ref{b}) with \emph{permuted\/}
quark-flavour structure of initial and final states, the proof of
our claim is a little bit more delicate. Also here, by the Landau
equations tetraquark-phile Feynman diagrams start contributing at
order $O(\alpha_{\rm s}^2)$. Quark--hadron duality
(Fig.~\ref{ESRR}) is established by noting that, at hadron level,
all members~of~the set of tetraquark-phile contributions exhibit
two-meson intermediate states, whereas all others don't, for both
the two-point correlators (\ref{2p}) (Fig.~\ref{2R}) and the
three-point correlators (\ref{3p}) (Fig.~\ref{3R}),
$$\left\langle{\rm T}\left(\theta_{\overline ad\overline cb}(y)\,
\theta^\dag_{\overline ab\overline cd}(x)\right)\right\rangle
\qquad\mbox{and}\qquad\left\langle{\rm T}\left(j_{\overline ad}(y)
\,j_{\overline cb}(y')\,\theta^\dag_{\overline ab\overline cd}(x)
\right)\right\rangle.$$

Using our tetraquark-phile spectral densities
$\rho_{\rm r},\Delta_{\rm r}$, and Fourier transforms $A(T\to
j_{\overline ad}\,j_{\overline cb})$ of $\langle0|{\rm T}
[j_{\overline ad}(y)\,j_{\overline cb}(y')]|T\rangle$, we find, as
flavour-\underline{r}eshuffling \emph{tetraquark-adequate\/} QCD
sum rules,

\begin{align*}f_{\overline ab\overline cd}\,f_{\overline
ad\overline cb}\exp(-M^2\,\tau)&=\hspace{-5.6ex}
\int\limits_{(m_a+m_b+m_c+m_d)^2}^{s_{\rm eff}(\tau)}
\hspace{-5.1ex}{\rm d}s\exp(-s\,\tau)\,\rho_{\rm r}(s)+
\mbox{Borelized power corrections}\ ,\\[1ex]f_{\overline
ab\overline cd}\,A(T\to j_{\overline ad}\,j_{\overline
cb})\exp(-M^2\,\tau)&=\hspace{-5.6ex}
\int\limits_{(m_a+m_b+m_c+m_d)^2}^{s_{\rm eff}(\tau)}
\hspace{-5.1ex}{\rm d}s\exp(-s\,\tau)\,\Delta_{\rm r}(s)+
\mbox{Borelized power corrections}\ .\end{align*}

\vspace{3.05445ex}\noindent\small{\bf Acknowledgements.} D.M.\ is
grateful for support by the Austrian Science Fund (FWF), Project
P29028. D.M.\ and H.S.\ express gratitude for support under joint
CNRS/RFBR Project PRC Russia/19-52-15022.\end{document}